\documentclass[amsmath,amssymb,twocolumn]{revtex4} 
\usepackage[parfill]{parskip}    
\usepackage{graphicx}
\usepackage{amssymb}
\usepackage{epstopdf}
\usepackage{color}
\usepackage{pdfpages}
\usepackage{soul}
\usepackage{amsmath}

\begin{document}
\title{Relativistic Anisotropic Fluid Spheres Satisfying a Non-Linear Equation of State}

\author{Francisco Tello-Ortiz}
\email{francisco.tello@ua.cl
 } \affiliation{Departamento de F\'{i}sica, Facultad de ciencias b\'{a}sicas,
Universidad de Antofagasta, Casilla 170, Antofagasta, Chile.}

\author{M. Malaver}
\email{mmalaver@umc.edu.ve}
\affiliation{Bijective Physics Institute, Idrija, Slovenia.}
\affiliation{Maritime University of the Caribbean, Department of Basic Sciences, Catia la Mar,
Venezuela.}

\author{\'Angel Rinc\'on}
\email{angel.rincon@pucv.cl}
\affiliation{Instituto de F\'isica, Pontificia Universidad Cat\'olica de Valpara\'iso, Avenida Brasil 2950, Casilla 4059, Valpara\'iso, Chile.}

\author{Y. Gomez-Leyton}
\email{ygomez@ucn.cl}
\affiliation{Departamento de F\'isica, Universidad Cat\'olica del Norte, Av. Angamos 0610, Antofagasta,
Chile.}

\begin{abstract}
In this work, a spherically symmetric and static relativistic anisotropic fluid sphere solution of the Einstein field equations is provided. To build this particular model, we have imposed metric potential $e^{2\lambda(r)}$ and an equation of state. Specifically, the so-called modified generalized Chaplygin equation of state with $\omega=1$ and depending on two parameters, namely, $A$ and $B$. These ingredients close the problem, at least mathematically. However, to check the feasibility of the model, a complete physical analysis has been performed. Thus, we analyze the obtained geometry and the main physical observables, such as the density $\rho$, the radial $p_{r}$, and tangential $p_{t}$ pressures as well as the anisotropy factor $\Delta$. Besides, the stability of the system has been checked by means of the velocities of the pressure waves and the relativistic adiabatic index. It is found that the configuration is stable in considering the adiabatic index criteria and is under hydrostatic balance. Finally, to mimic a realistic compact object, we have imposed the radius to be $R=9.5\ [km]$. With this information and taking different values of the parameter $A$ the total mass of the object has been determined. The resulting numerical values for the principal variables of the model established that the structure could represent a quark (strange) star mixed with dark energy.      
\end{abstract}

\maketitle

\section{Introduction}
The highly non-linear nature of Einstein's field equations makes its solution a great challenge. One way to reduce such complexity in the context of stellar interiors, for example, is to set the space-time metric to be static and spherically symmetric. The latter decode the problem of second-order partial differential equations into ordinary second-order differential equations. Although the equations remain strongly coupled, the problem is greatly simplified. However, depending on the global ingredients of which the matter distribution of the fluid sphere is composed, that is, isotropic, anisotropic content with or without the inclusion of an electric charge (which will make the situation more complicated) solve the resulting equations is not simple. Among of all possibilities to solve the equations in an analytically way, the simplest is to assume that the material content is isotropic $p_ {r}=p_ {t}$ (without the inclusion of electric charge) \cite{r1}. Since in this case the problem contains only four variables to be determined, namely the energy-density $\rho$, the isotropic pressure $p$, and both metric potentials $e^{2\nu}$ and $e^{2\lambda}$. In this case, it is inevitable (although partially) to assume the form of one of the metric potentials (usually $e^{2\lambda}$), from which, together with the isotropy condition, the remaining variables are determined by closing the problem. On the other hand, the description of a more realistic situation requires the inclusion of local anisotropy \cite{r2,r3}. However, this means that the number of variables to be determined is five (in the case of anisotropic distributions with an electric charge, the number rises to six), then it is necessary to prescribe more information. In this sense, an interesting prescription is the imposition of an equation of state (EoS). In this regard, from a more phenomenological perspective, this relationship that links the main thermodynamic functions of the fluid \i.e, the energy-density and pressure (in the radial and tangential directions), and that describes the microphysical processes of the system, is estimated to be a linear relationship between those physical observables \cite{r4}. The usage of EoS to recreate the behavior of the matter content inside the stellar interior has been used as a building block to obtain admissible models that, in principle, could represent compact structures such as neutron or quark stars. In this respect, some works available in the literature have addressed the study of compact structures within the framework of general relativity by using linear EoS \cite{r5,mala6,r6b}, non-linear EoS such as Van der Waal \cite{mala7, thi,erra, r6} or the color superconductivity EoS \cite{r7}. Moreover, the well known MIT EoS \cite{r8,r9}  which corresponds to a special case of the color superconductivity EoS \cite{r10,r11,r12} has been used in the context of $f(R,\mathcal{T})$ \cite{r13} to investigate the possibilities of obtaining ultra-high dense compact objects describing quark stars.

On the other hand, general relativity not only addresses the study of stellar interiors.  Another critical issue to study within the scope of Einstein gravity theory is, for example, the cosmological scenario. This vast area contemplates the problem of the existence of dark components in the Universe \i.e, dark matter, and dark energy. In this regard, the so-called dark fluids \cite{r14}, such as Phantom \cite{r15} and Quintessence \cite{r16} fields, to name a few, were introduced to explain the accelerated expansion of the Universe. As an alternative to the Phantom and Quintessence fields is the so-called Chaplygin gas \cite{r17,r18}. The following EoS drives this peculiar and intriguing fluid
\begin{equation}\label{eq1}
p=-\frac{B}{\rho},    
\end{equation}
where $p$ is the pressure, $\rho$ the energy-density and $B$ a positive constant with units of $\text{length}^{-4}$ (using geometrized relativistic units where $8\pi G=c=1$). To fit within the observational data the above equation (\ref{eq1}) was generalized \cite{r19} as follows
\begin{equation}\label{eq2}
p=-\frac{B}{\rho^{\omega}},    
\end{equation}
where the parameter $\omega$ is restricted to belong to $(0,1]$. Of course, the case $\omega=1$ leads to Eq. (\ref{eq1}). This generalized version of the Chaplygin EoS was studied under the presence of viscosity \cite{r20,r21}. Besides, an extended version of the generalized Chaplygin EoS was provided in \cite{r22}. Specifically, this extension concerns in the sum of two parts: i) the generalized EoS plus ii) a linear term in $\rho$. Explicitly it reads
\begin{equation}\label{eq3}
p=A\rho-\frac{B}{\rho^{\omega}},
\end{equation}
being $A$ a positive parameter constrained to $0<A<1/3$. This extension can be re-extended to include barotropic fluid EoS \cite{r23,r24}. Chaplygin EoS has been extensively used in different context. For example in the construction of charged anisotropic fluid spheres \cite{r25}, compact structures in the arena of $f(T)$ gravity theory \cite{r26,r27}, wormhole geometries \cite{r28} and $5$-dimensional cosmology \cite{r29} to name a few. As can be seen, Chaplygin gas and its extensions or generalizations have proven to be versatile tools to face many open problems at the theoretical level. Following the same direction the spirit of this article concerns in obtaining new analytical relativistic anisotropic fluid spheres where the main thermodynamic variables are linked via the EoS (\ref{eq3}). Furthermore, to close the problem we have imposed the $e^{2\lambda(r)}$ metric potential previously used in \cite{r30}. To produce a more realistic situation, we have imposed the radius of the compact object to be $R=9.5\ [km]$. This value is within the range of the reported observational data for some compact structures \cite{r31}. The feasibility of our model is tested by means of the established criteria and statements for anisotropic matter distributions in the arena of general relativity \cite{r32,r33,r34,r35,r36,r37,r38,r39,r40,r41,r42,r43,r44,r45,r46,r47,r48,r49,r50,r51,r52,r53,r54,r55}. The resulting model can be compared with the abundant works available in the literature and contrasted with astrophysical phenomenology \cite{r56,r57,r58,r59,r60,r61,r62,payel,r63,r64,r65,r66,r67,ili,r68,r69,r70} (and references contained therein). 

The article is organized as follows: In Sec. \ref{sec2} the general field equations for anisotropic matter distributions are presented. Sec. \ref{sec3} presents the model, exhibiting the essential ingredients, such as the geometry and thermodynamic description. In Sec. \ref{sec4}, a complete mathematical and physical analysis is performed, studying the most relevant aspect of the solution. Sec. \ref{sec5} talks about the Israel-Darmois junction mechanism with the vacuum space-time described by the well known Schwarzschild solution. The matching condition procedure allows determining the full set of the constant parameter that characterize the model as well as the total mass contained by the fluid sphere. In Sec. \ref{sec6} is realized the balance and stability analysis. To achieve it we have employed the generalized or modified Tolman-Oppenheimer-Volkoff equation and the relativistic adiabatic index and Abreu's criterion. Sec. \ref{sec7} refers to the generating function of the present model. Finally, Sec. \ref{sec8} concludes the work.

Throughout the study we shall employ the mostly positive signature $\{-,+,+,+\}$ and units where $8\pi G=c=1$ then $\kappa\equiv 8\pi G/c^{4}=1$. Besides, to solve the mathematical part of this article we have employed Maple 17 programming.  

\section{General Field Equations} \label{sec2}
To describe a spherically symmetric and static compact structure, one has the most general line element given by
\begin{equation}\label{eq4}
ds^{2}=-e^{2\nu(r)}dt^{2}+e^{2\lambda(r)}dr^{2}+r^{2}\left(d\theta^{2}+\text{sen}^{2}\theta d\phi^{2}\right)    
\end{equation}
in canonical coordinates $\{x^{\alpha}\}=\{x^{0},x^{1},x^{2},x^{3}\}=\{t,r,\theta,\phi\}$. The matter distribution of this stellar interior is characterized by the following energy-momentum tensor 
\begin{equation}\label{eq5}
T_{\mu\nu}=\text{diag}\left(-\rho,p_{r},p_{t},p_{t}\right).   
\end{equation}
As can be seen, Eq. (\ref{eq5}) represents an anisotropic fluid distribution \i.e, $p_{r}\neq p_{t}$. Therefore the local anisotropies are measure by the anisotropy factor defined as $\Delta\equiv p_{t}-p_{r}$. The main physical variables that characterize this imperfect fluid are the density $\rho$, the radial pressure $p_{r}$ and the tangential pressure $p_{t}$. The radial and tangential pressures are measured relative to the comoving fluid $4$-velocity $u^{\alpha}=e^{-\nu(r)}\delta^{\alpha}_{\ 0}$. The field equations describing the gravitational interaction are given by Einstein field equations
\begin{equation}\label{eq6}
G_{\mu\nu}\equiv R_{\mu\nu}-\frac{1}{2}Rg_{\mu\nu}=\kappa T_{\mu\nu}.    
\end{equation}
Next, putting together Eqs. (\ref{eq4}), (\ref{eq5}) and (\ref{eq6}) one arrives at the following set of equations 
\begin{eqnarray}\label{eq7}
\frac{1}{r^{2}}\left[r\left(1-e^{-2\lambda}\right)\right]^{\prime}&=&\rho, \\ \label{eq8}
-\frac{1}{r^{2}}\left(1-e^{-2\lambda}\right)+\frac{2\nu^{\prime}}{r}e^{-2\lambda}&=&p_{r}, \\ \label{eq9}
e^{-2\lambda}\left(\nu^{\prime \prime}+\nu^{\prime 2}+\frac{\nu^{\prime}}{r}-\nu^{\prime}\lambda^{\prime}-\frac{\lambda^{\prime}}{r}\right)&=&p_{t},
\end{eqnarray}
where primes denote differentiation w.r.t. the radial coordinate $r$. As it is well known, solve the system of equations (\ref{eq7})-(\ref{eq9}) is not a simple task. Nevertheless, one can re-express the above set of equations to minimize the mathematical complications by introducing the so-called Durgapal-Bannerji transformation \cite{r71} 
\begin{equation}\label{r10}
x=r^{2}, \quad Z(x)=e^{-2\lambda(r)}, \quad D^{2}y^{2}(x)=e^{2\nu(r)},   
\end{equation}
being $D$ an arbitrary constant.
Thus, the system (\ref{eq7})-(\ref{eq9}) becomes \begin{eqnarray}\label{eq11}
\frac{1-Z}{x}-2\dot{Z}&=&\rho, \\ \label{eq12}
4Z\frac{\dot{y}}{y}+\frac{Z-1}{x}&=&p_{r}, \\ \label{eq13}
4xZ\frac{\ddot{y}}{y}+\left(4Z+2x\dot{Z}\right)\frac{\dot{y}}{y}+\dot{Z}&=&p_{t},
\end{eqnarray}
where dots mean differentiation w.r.t. the $x$ variable. With this redefinition in hand, the mass contained within the sphere is given by
\begin{equation}\label{eq14}
m(x)=\frac{1}{4}\int^{x}_{0}\sqrt{\tilde{x}}\rho(\tilde{x})d\tilde{x}.    
\end{equation}
In the search for analytical solutions of the Einstein field equations, it is necessary to prescribe certain information. Within the possibilities is to specify the complete geometry that describes the stellar interior, that is, to impose the metric potentials $e^{2\lambda}$ and $e^{2\nu}$ or to impose one metric potential and link the density $\rho$ with the pressure (usually with the radial pressure). On this occasion, we will specify the metric potential $e^{2\lambda}$ and impose an EoS, specifically Eq. (\ref{eq3}) with $\omega=1$. In general, one has
\begin{equation}\label{eq15}
p_{r}=p_{r}(\rho).   
\end{equation}
So, combining Eqs. (\ref{eq11}), (\ref{eq12}) and (\ref{eq15}) one obtains \cite{r5,r6}
\begin{equation}\label{eq16}
y=dx^{-\frac{1}{4}}\text{Exp}\left[\int \frac{1+xp_{r}(\rho)}{4xZ}dx\right].  
\end{equation}
Then, the line element representing the stellar interior can be written in general form as follow
\begin{equation}\label{eq17}
\begin{split}
ds^{2}=-d^{2}r^{-1}\text{Exp}\left[\int \left(\frac{1+xp_{r}(\rho)}{4xZ}\right)dr\right]dt^{2}+Z^{-1}dr^{2} &\\ +r^{2}\left(d\theta^{2}+\text{sen}^{2}\theta d\phi^{2}\right).
\end{split}
\end{equation}
Therefore, with this general description of the stellar interior in hand, by fixing one metric potential and providing an EoS ($p_{r}=p_{r}(\rho)$), the problem is closed from the mathematical point of view.

\section{The Model: Modified Chaplygin EoS} \label{sec3}
As stated in the previous section \ref{sec2}, to close the problem at least from the mathematical point of view, one needs to supplement the problem with extra information. In this opportunity following \cite{r30} we impose the following $e^{2\lambda}$ metric potential
\begin{equation}\label{eq18}
e^{2\lambda(r)}=\left(\frac{1+Cr^{2}}{1-Cr^{2}}\right)^{n},   
\end{equation}
where $n$ is a positive integer number and $C$ a constant parameter with units of $\text{length}^{-2}$. The choice (\ref{eq18}) is well motivated because is free from mathematical singularities at $r=0$ (the center of the object), specifically $e^{2\lambda(r)}|_{r=0}=1$. Moreover as we will see (\ref{eq18}) yields to a well behaved density function $\rho(r)$ and in consequence a well established mass function $m(r)$. So, re-written Eq. (\ref{eq18}) in the language of Durgapal-Bannerji transformation one obtains
\begin{equation}\label{eq19}
Z(x)=\left(\frac{1-Cx}{1+Cx}\right)^{n}.  
\end{equation}
From now on, we shall assume throughout the work $n=1$. Next, inserting (\ref{eq19}) into Eq. (\ref{eq11}) one arrives to
\begin{equation}\label{eq20}
\rho(x)=\frac{2C\left(Cx+3\right)}{\left(1+Cx\right)^{2}}.   
\end{equation}
Now from Eqs. (\ref{eq14}) and (\ref{eq20}) one gets the following mass function 
\begin{equation}\label{eq21}
m(x)=\frac{Cx^{3/2}}{1+Cx}.    
\end{equation}
To complete the geometric and thermodynamic description of the model we will use the generalized Chaplygin EoS given by Eq. (\ref{eq3}) along with $\omega=1$, that is
\begin{equation}\label{chaplygin}
p_{r}=A\rho-\frac{B}{\rho}. 
\end{equation}
So, putting together Eqs. (\ref{eq3}), (\ref{eq16}), (\ref{eq19}) and (\ref{eq20}) one obtains
\begin{equation}\label{eq22}
\begin{split}
y(x)=E\left(Cx+3\right)^{\frac{B}{4C^{2}}}\left(Cx+1\right)^{\frac{A}{2}}\left(Cx-1\right)^{-A-\frac{1}{2}+\frac{B}{4C^{2}}} & \\ \times \text{Exp}\left[\frac{Bx\left(Cx+2\right)}{16C}\right].
\end{split}
\end{equation}
Next, the expressions for the radial $p_{r}(x)$ and transverse $p_{t}(x)$ pressures are given by 
\begin{equation}\label{eq23}
\begin{split}
p_{r}(x)=\frac{1}{2C\left(Cx+3\right)\left(Cx+1\right)^{2}}\bigg[-BC^{4}x^{4}+4AC^{4}x^{2} &\\-4BC^{3}x^{3}+24AC^{3}x-6BC^{2}x^{2}+36AC^{2}-4BCx-B\bigg],
\end{split}
\end{equation}
and
\begin{equation}\label{eq24}
\begin{split}
p_{t}(x)=\frac{-1}{16C^{2}\left(Cx+3\right)^{2}\left(Cx+1\right)^{3}}\bigg[B^{2}C^{7}x^{8}+9B^{2}C^{6}x^{7}&\\
+8ABC^{7}x^{6}+16BC^{7}x^{6}+33B^{2}C^{5}x^{6}+56ABC^{6}x^{5}&\\+16A^{2}C^{7}x^{4}+128BC^{6}x^{5}+57B^{2}C^{4}x^{5}+128ABC^{5}x^{4}&\\
+80A^{2}C^{6}x^{3}+344BC^{5}x^{4}+64AC^{6}x^{2}+31B^{2}C^{3}x^{4}&\\
+64ABC^{4}x^{3}+32C^{6}x^{3}+48A^{2}C^{5}x^{2}+384BC^{4}x^{3}
&\\ +352AC^{5}x^{2}-41B^{2}C^{2}x^{3}-136ABC^{3}x^{2}+224C^{5}x^{2}&\\
-144A^{2}C^{2}x+80BC^{3}x^{2}+384AC^{4}x-65B^{2}Cx^{2}&\\
-120ABC^{2}x+480C^{4}x-192BC^{2}x-288AC^{3}&\\
-25B^{2}x+288C^{3}-120BC\bigg].
\end{split}
\end{equation}
At this point the full geometric and thermodynamic description of the model is given by Eqs. (\ref{eq19}), (\ref{eq20}), (\ref{eq22}), (\ref{eq23}) and (\ref{eq24}). In the next section, we will provided a complete physical and mathematical analysis of the present model.  

\section{Geometric and Thermodynamic Analysis}\label{sec4}
\begin{figure}
\includegraphics[width=7.5cm,height=5.5cm]{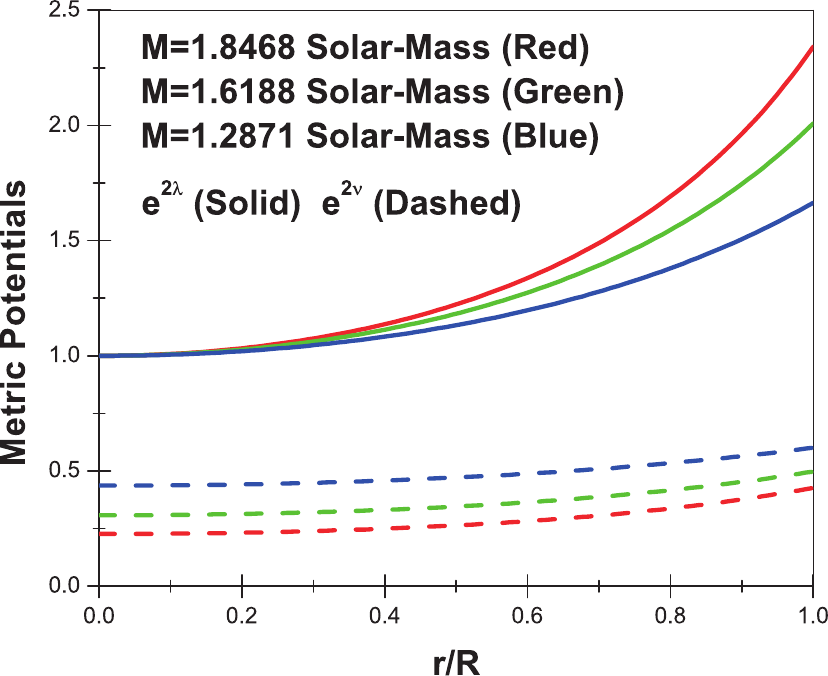}
\caption{Variation of metric potentials with the dimensionless radial coordinate $r/R$ with $R=9.5\ [km]$ and for different values of the parameters $A$, $B$, $C$ and $F$ mentioned in table \ref{table1}.}
\label{fig1}
\end{figure}

At this level, we should notice that the transformation made should be invested. The latter means that we now need to map our solution from $x$ to the original radial coordinate $r$. Just after such point is when an appropriated physical and mathematical analysis on the geometry can be achieved.
%
%
Thus, the the metric potentials describing the geometry of the stellar interior are
\begin{equation}\label{eq26}
e^{2\lambda(r)}=\frac{1+Cr^{2}}{1-Cr^{2}},   
\end{equation}
\begin{equation}\label{eq27}
\begin{split}
e^{2\nu(r)}=F^{2}\left(Cr^{2}+3\right)^{\frac{B}{2C^{2}}}\left(Cr^{2}+1\right)^{A}\left(Cr^{2}-1\right)^{-2A-1+\frac{B}{2C^{2}}}& \\ \times \text{Exp}\left[\frac{Br^{2}\left(Cr^{2}+2\right)}{8C}\right],
\end{split}
\end{equation}
where $F^{2}\equiv D^{2}E^{2}$. As it is observed, the radial component of the metric tensor (\ref{eq26}) is completely regular for all $r$ belonging to $[0,R]$ where $R$ defines the size (the radius) of the structure. Specifically, at $r=0$, one has $e^{2\lambda(r)}|_{r=0}=1$. Moreover, the metric potential (\ref{eq26}) is an increasing function with increasing $r$ at every point belonging to the interval $[0,R]$. Regarding the temporal component of the metric potential given by (\ref{eq27}) to ensure a well behaved stellar interior without pathologies, we have restricted the exponent of the term $\left(Cr^{2}-1\right)$ in (\ref{eq27}) to
\begin{equation}\label{eq28}
-2A-1+\frac{B}{2C^{2}}=k,    
\end{equation}
being $k$ an integer number. In this concern, we have considered the simplest choice taken $k=0$. The reasons for the constraint (\ref{eq28}) are clear. In fact, at $r=0$ from Eq. (\ref{eq27}) one obtains
\begin{equation}\label{eq29}
e^{2\nu(r)}|_{r=0}=F^{2}\left(3\right)^{\frac{B}{2C^{2}}}\left(-1\right)^{-2A-1+\frac{B}{2C^{2}}}.   
\end{equation}
So, if (\ref{eq28}) is not a integer number, then we could get a complex number at the center of the object. Therefore, in order to avoid this problem it is necessary to impose (\ref{eq28}). Besides, this guarantees $e^{2\nu(r)}|_{r=0}>0$ and $\left(e^{2\nu(r)}\right)^{\prime}|_{r=0}=0$. With the previous assumption Eq. (\ref{eq27}) becomes
\begin{equation}\label{eq30}
\begin{split}
e^{2\nu(r)}=F^{2}\left(Cr^{2}+3\right)^{\frac{B}{2C^{2}}}\left(Cr^{2}+1\right)^{A}  \text{Exp}\left[\frac{Br^{2}\left(Cr^{2}+2\right)}{8C}\right].
\end{split}
\end{equation}
Finally, the stellar interior geometry is represented by Eqs. (\ref{eq26}) and (\ref{eq30}). The trend of both metric potentials is depicted in Fig. \ref{fig1}. As it is observed, both are monotone increasing functions from the center of the object $r=0$ towards the boundary $r=R$. 

\begin{figure*}[ht]
\centering
\includegraphics[width=0.32\textwidth]{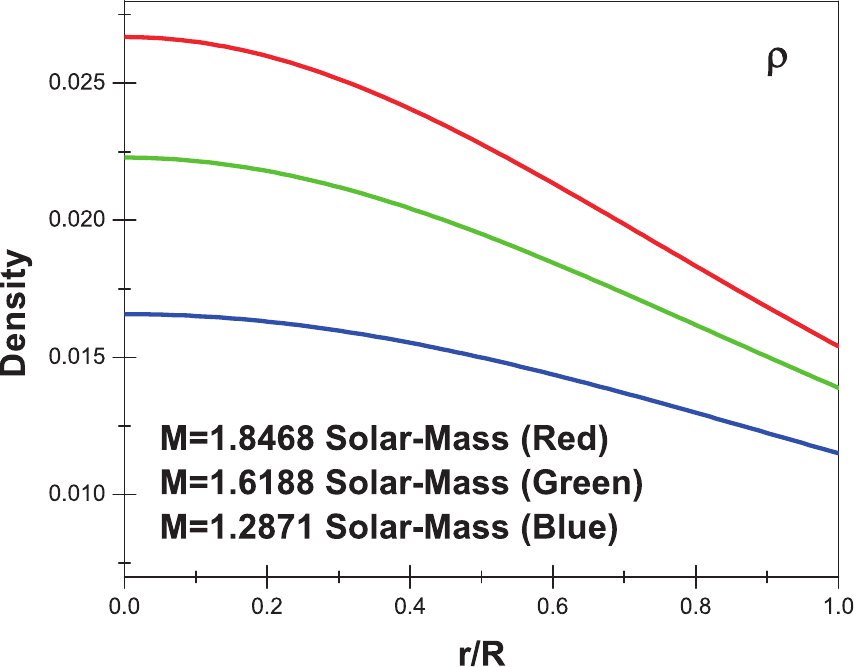}     \
\includegraphics[width=0.32\textwidth]{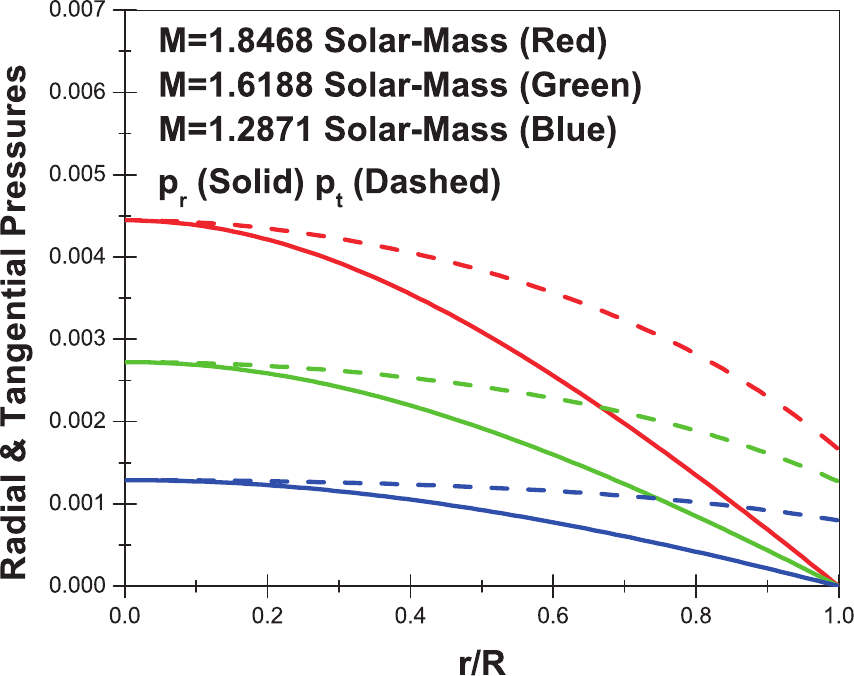}      \
\includegraphics[width=0.32\textwidth]{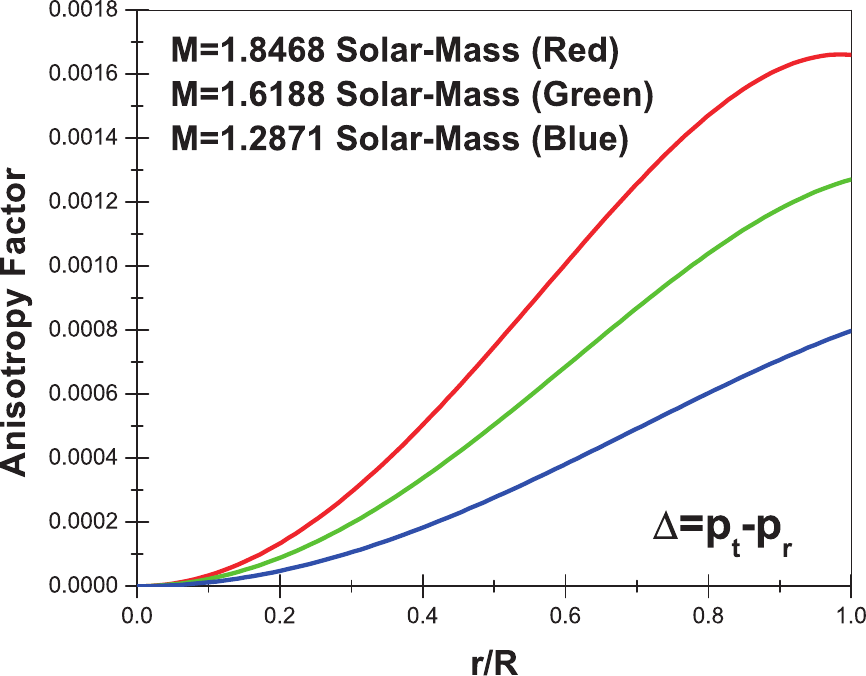}     \
\caption{
{\bf Left Panel}: The density $\rho$ profile against the dimensionless radial coordinate $r/R$. {\bf Middle Panel}: The radial $p_{r}$ and tangential $p_{t}$ pressures versus the dimensionless radial coordinate $r/R$.  {\bf Right Panel}: The anisotropy factor $\Delta$ versus $r/R$. All these plots have been built 
by using $R=9.5\ [km]$ and different values of the parameters $A$, $B$, $C$ and $F$ mentioned in table \ref{table1}. It should be noted that, as we are working in relativistic geometrized units, the vertical axis has units of $[km^{-2}]$ for all panels.
}
\label{fig2}
\end{figure*}

As the field equations dictate, the curvature encoded and described by the geometry of the space-time is related to the matter content. So, an appropriated geometry leads to a well behaved thermodynamic variables driven the matter field inside the compact structure. In this regard, the election of (\ref{eq26}) yields to a well defined and positive density 
\begin{equation}\label{eq31}
\rho(r)=\frac{2C\left(Cr^{2}+3\right)}{\left(Cr^{2}+1\right)^{2}}    
\end{equation}
at every point within the compact object. The left panel of Fig. \ref{fig2} displays the behavior of the density throughout the stellar interior. 
As can be seen, this primary quantity gets its maximum value at the center of the object and its minimum at the surface.
%
%
What is more, as the mass increases, the object becomes denser at the core (as shows the red curve). On the other hand, the central density $\rho(0)=\rho_{c}$ provides
\begin{equation}\label{eq32}
\rho_{c}=6C.    
\end{equation}
Then, to assure a positive defined density everywhere, the parameter $C$ must be positive. This fact is confirmed by table \ref{table1}, where for different choices of the parameter $A$ the constant $C$ is positive. Moreover, in table \ref{table2} are depicted the central $\rho_{c}$ and surface $\rho_{s}$ ($\rho_{s}=\rho(R)$) density values for all cases. As can be seen, the central and surface density are above the nuclear density saturation ($2.8\times 10^{14}\ [g/cm^{3}]$) in the cases $A=0.25$ and $A=0.2$. As the $A$ parameter decreases in magnitude both, the central and surface density decrease. Moreover, for the case $A=0.15$, the surface density is below the nuclear density saturation. Also, as mentioned before, both the central and the surface density increase in magnitude as the total mass increases. Regarding the pressure waves in the main direction of the structure \i.e, the radial and transverse ones. The middle panel of Fig. \ref{fig2} exhibits the trend of both quantities. As it is required $p_{r}$ and $p_{t}$ are decreasing functions at all points inside the stellar interior. Furthermore, the tangential pressure $p_{t}$ dominates the radial pressure $p_{r}$ everywhere. The fact $p_{t}>p_{r}$ implies that $\Delta\equiv p_{t}-p_{r}>0$. The quantity $\Delta$ is the so-called anisotropy factor, it measures the difference between the radial and tangential pressures inside the stellar interior and realizes of local anisotropies introduced by the imperfect matter fluid distribution under consideration (\ref{eq5}). The role played by $\Delta$ is very important in studying compact structures within the arena of general relativity. The anisotropy factor $\Delta$ has a direct incidence in the production of more compact objects \cite{r45} and in the stability and balance mechanism (as we will see later). Regarding the latest, the presence of $\Delta$ may or may not be favorable for the system. In this respect, a positive $\Delta$ implies the presence of an additional force repulsive in nature into the configuration. Otherwise, if $\Delta<0$, which means by definition that the radial pressure dominates the tangential one, the new force is attractive in nature. This fact is not favorable for the system due to the object could, in principle, collapse onto a point singularity. The trend of the anisotropy factor $\Delta$ is illustrated in the right panel of Fig. \ref{fig2}. As was discussed $\Delta>0$ everywhere, reaching its maximum values at the surface of the star while at the center is zero. Indeed, due to the spherical symmetry $p_{r}(0)=p_{t}(0)$ then $\Delta(0)=0$. From the expressions (\ref{eq23}) and (\ref{eq24}) at $r=0$ one obtains 
\begin{equation}\label{eq33}
p_{r}(0)=\frac{36AC^{2}-B}{6C},
\end{equation}
and
\begin{equation}\label{eq34}
p_{t}(0)=\frac{12AC^{2}-12C^{2}+5B}{6C}.
\end{equation}
So, to ensure $\Delta(0)=0$ we equate Eqs. (\ref{eq33}) and (\ref{eq34}) arriving to
\begin{equation}\label{eq35}
-2A-1+\frac{B}{2C^{2}}=0.    
\end{equation}
The expression (\ref{eq35}) coincides with (\ref{eq28}) when $k=0$. This confirms that the only possibility to get the condition $\Delta(0)=0$ is taking $k=0$. In table \ref{table2} are placed the values of the central pressure. As can be seen, the central pressure decreases as the parameter $A$ decreases in magnitude.      

\begin{figure*}[ht]
\centering
\includegraphics[width=0.32\textwidth]{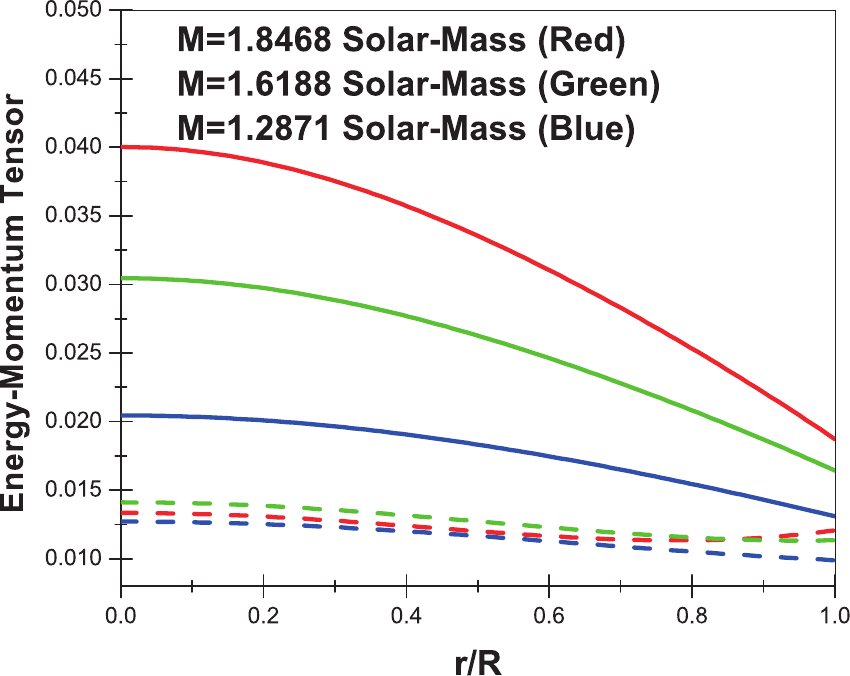}     \
\includegraphics[width=0.32\textwidth]{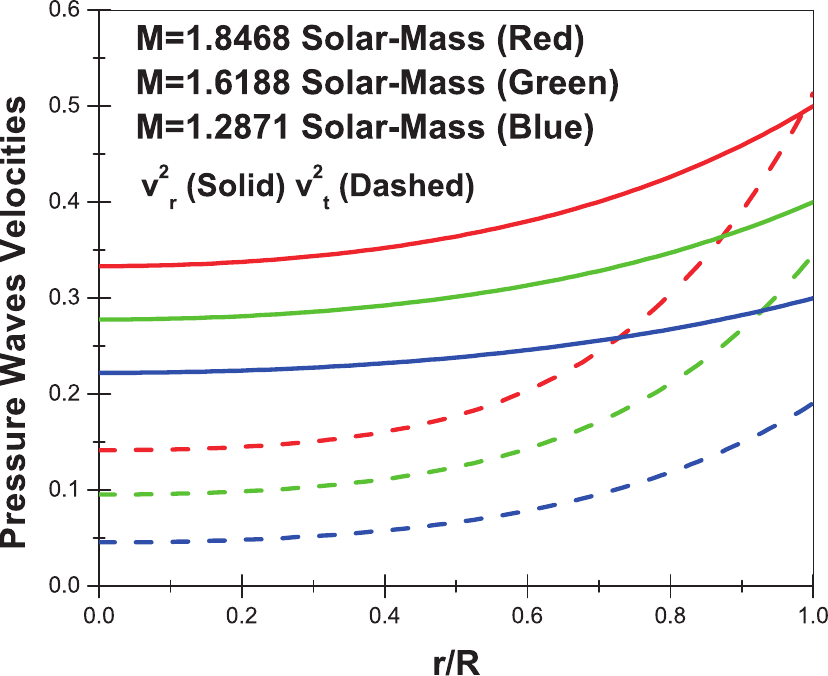}      \
\caption{
{\bf Left Panel}: The energy-momentum tensor against the dimensionless radial coordinate $r/R$. Solid lines correspond to $\rho+p_{r}+2p_{t}$ and dashed lines to $\rho-p_{r}-2p_{t}$. {\bf Right Panel}: The pressure waves velocities versus $r/R$. All these plots have been built by using $R=9.5\ [km]$ and different values of the parameters $A$, $B$, $C$ and $F$ mentioned in table \ref{table1}. It should be noted that, as we are working in relativistic geometrized units, the vertical axis has units of $[km^{-2}]$ for the left panel, while for the right one is dimensionless.
}
\label{fig3}
\end{figure*}

With the thermodynamic description in hand, one can infer in broad strokes the behavior of the matter content in the stellar interior. This is done by analyzing the conditions that the energy-momentum tensor must meet \cite{r51}
\begin{equation}\label{eq36}
\rho+p_{r}+2p_{t}\geq 0 \quad \mbox{and} \quad  \rho-p_{r}-2p_{t}\geq 0.     
\end{equation}
The inequalities given in (\ref{eq36}) say that to have a well-defined energy-momentum tensor describing the matter field, they must be satisfied simultaneously. This fact entails that the matter distribution inside the start corresponds to a standard matter content. The left panel in Fig. \ref{fig3} shows that both inequalities are satisfied at every point inside the star. The solid curves correspond to the left inequality in (\ref{eq36}), and dashed lines represent the right inequality in (\ref{eq36}). Moreover, the behavior of the matter content also is subject to causality condition. This condition says that the sound speeds of the pressure waves in the principal directions of the fluid sphere do not exceed the speed of light ($c=1$ in relativistic geometrized units). The sound velocities in the radial and transverse directions of the structure are defined as follows
\begin{equation}\label{eq37}
v^{2}_{r}(r)=\frac{dp_{r}(r)}{d\rho(r)} \quad \mbox{and} \quad  v^{2}_{t}(r)=\frac{dp_{t}(r)}{d\rho(r)}. 
\end{equation}
So, to satisfy causality condition both expressions must be restricted to: $0\leq v^{2}_{r}\leq 1$ and $0\leq v^{2}_{t}\leq 1$. As the right panel of Fig. \ref{fig3} corroborates $v^{2}_{r}$ and $v^{2}_{t}$ are less than the speed of light everywhere inside the structure. It is worth mentioning that the velocities associated with fluid pressure waves are not necessarily decreasing when there is anisotropy in the stellar medium, in distinction with what happens with its isotropic counterpart. 
An important fact to be noted here, is that as the parameter $A$ increases in magnitude, at some point within the stellar interior (at $r/R=0.98$ approximately) the transverse sound speed $v^{2}_{t}$ overcomes the radial sound speed $v^{2}_{r}$ of the pressure waves (see the red curves in the right panel of Fig. \ref{fig3}). This issue as we will see later has a direct incidence on the stability of the structure. 

\section{Israel-Darmois Junction Conditions}\label{sec5}
In the study of stellar interiors, a crucial point is the junction condition process. The matching condition mechanism allows determining the full set of constant parameters that describe the model and the macro observables such as the radius $R$ and the total mass $M$ of the fluid sphere. To carry out this process, one employs the well known Israel-Darmois junction conditions \cite{r72,r73}. As we are dealing with an uncharged and static configuration, the model is surrounded by an empty space-time described by the Schwarzschild solution \cite{r74} given by
\begin{equation}\label{eq38}
ds^{2}=-\left(1-\frac{2M_{\text{Sch}}}{r}\right)dt^{2}+\left(1-\frac{2M_{\text{Sch}}}{r}\right)^{-1}dr^{2}+r^{2}d\Omega^{2},\end{equation}
being $M_{\text{Sch}}$ the Schwarzschild mass. So, at the boundary of the object $\Sigma\equiv r=R$ the Israel-Darmois matching conditions dictate: i) the inner $\mathcal{M}^{-}$ and outer $\mathcal{M}^{+}$ manifolds given by (\ref{eq26}) and (\ref{eq30}) and (\ref{eq38}) respectively, induce on $\Sigma$ an intrinsic geometry represented by the metric tensor $g_{\mu\nu}$. The continuity of the metric tensor components across $\Sigma$ constitutes the first fundamental form and it means that $[ds^{2}]_{\Sigma}=0$. ii) Also is induced by $\mathcal{M}^{-}$ and $\mathcal{M}^{+}$ on $\Sigma$ an extrinsic geometry described by the extrinsic curvature tensor $K_{ij}$ (where $i$ and $j$ run over $\{x^{1},x^{2},x^{3}\}=\{r,\theta,\phi\}$). The continuity of the component $r-r$ of the extrinsic curvature tensor across the boundary of the object implies the second fundamental form which reads $p_{r}(r)|_{\Sigma}=0$, and the continuity of the $\theta-\theta$ and $\phi-\phi$ components leads to determine the total mass inside the sphere $m(R)=M$. At this stage, some comments are pertinent. First, to obtain the total mass contained by the fluid sphere, there are at least three equivalent ways: i) integrating the density $\rho$ in the interval $[0,R]$, ii) by imposing the continuity of the radial components of the metric tensor $g_{\mu\nu}$ across $\Sigma$ and iii) from the continuity of the angular components of the extrinsic curvature tensor. Second, a vanishing radial pressure at the surface of the configuration determines the size of the object \i.e, its radius $R$, and also avoids an indefinite expansion of the system confined the matter field between $r=0$ and $r=R$. So, explicitly the first fundamental form reads      
\begin{equation}\label{eq39}
\frac{1-CR^{2}}{1+CR^{2}}=1-\frac{2M}{R},    
\end{equation}
leading to 
\begin{equation}\label{eq40}
M=\frac{CR^{3}}{1+CR^{2}},    
\end{equation}
and 
\begin{equation}\label{eq41}
\begin{split}
F^{2}\left(CR^{2}+3\right)^{\frac{B}{2C^{2}}}\left(CR^{2}+1\right)^{A}  \text{Exp}\left[\frac{BR^{2}\left(CR^{2}+2\right)}{16C}\right]&\\=1-\frac{2M}{R}.
\end{split}
\end{equation}
In the previous expression, $M$ corresponds to the total mass contained by the sphere, which at $\Sigma$ coincides with the Schwarzschild mass $M_{\text{Sch}}$. On the other hand, the second fundamental form entails    
\begin{equation}\label{eq42}
\begin{split}
p_{r}(R)=0 \Rightarrow 
B=\frac{4AC^{2}\left(3+CR^{2}\right)^{2}}{\left(1+CR^{2}\right)^{4}}.
\end{split}
\end{equation}
The condition (\ref{eq35}) and the expressions (\ref{eq40})-(\ref{eq42}) are the necessary conditions to determine all the parameters that characterized the model, namely $B$, $C$, $F$ and $M$ along with the imposition of the radius $R$ and the parameter $A$ constrained to $0<A<1/3$. All the numerical values obtained for the present study are reported in table \ref{table1} and the mass $M$ in table \ref{table3} for a fixed radius and different values of the constant $A$. From table \ref{table3} it is observed that the total mass $M$ increases with increasing $A$ (and vice versa). Moreover, regarding the central parameter exhibited in table \ref{table2} and the mass in table \ref{table3} we can conclude prematurely that the model represents a quark star whose principal observable components such as density and radial pressure describe a micro-physics dominated by dark energy. 

Nevertheless, to assure that the salient toy model could represent a real compact structure, it is important to highlight the role played by the arbitrary constant parameters that characterize the solution, namely $\{A,B,C,F\}$. In this regard, the most important parameters are $\{A,B,C\}$, this can be elucidated from Eqs. (\ref{eq35}), (\ref{eq40}) and (\ref{eq42}). The fact to restrict $A$ between 0 and 1/3 leads to the central density of the object to an order of magnitude greater than the  nuclear density saturation, specifically $ 10^{15}\ [g/cm^{3}]$ (or at least of the same order $ 10^{14}\ [g/cm^{3}]$). Then, by combining Eqs. (\ref{eq35}) and (\ref{eq42}) the parameters $B$ and $C$ can be determined after $A$ and $R$ are fixed. The output assures a vanishing radial pressure at the boundary $\Sigma$, this fact is relevant since if $p_{r}(R)\neq 0$ the object will expand indefinitely. Moreover, once $C$ is obtained, this parameter fix the total mass $M$ of the fluid sphere as can be seen from Eq. (\ref{eq40}). So, the reported values in tables \ref{table2}  and \ref{table3} are in complete agreement with previous results considering quark stars supported by realistic equations of state \cite{nature}. Finally, from Eq. (\ref{eq41}) the parameter $F$ is determined, en this case $F$ sets the value of the temporal metric potential at the center of the compact configuration.

\begin{table*}
\caption{The numerical values of constant parameters $B$, $C$ and $F$ for a fixed radius $R=9.5\ [{km}]$ and for different values of the parameter $A$. }
\label{table1}
\begin{tabular*}{\textwidth}{@{\extracolsep{\fill}}lrrrrrrrl@{}}
\hline
$$A$ (\text{dimensionless})$& \multicolumn{1}{c}{$B$$\times 10^{-5}$\ [${km}^{-4}$]} & \multicolumn{1}{c}{$C$$\times 10^{-3}$\ [$\text{km}^{-2}$]} & \multicolumn{1}{c}{$F$ (\text{dimensionless})}  \\
\hline
$0.25$& $5.932834033$&$4.447034230$ & $0.2088237772$  \\
$0.20$& $3.862315863$&$3.714025475$ & $0.2572462264$    \\
$0.15$& $1.985200479$&$2.763220306$ & $0.3234834214$\\
\hline
\end{tabular*}
\end{table*}

\begin{table*}
\caption{The numerical values of some thermodynamic observables for a fixed radius $R=9.5\ [{km}]$ and for different values of the parameter $A$. }
\label{table2}
\begin{tabular*}{\textwidth}{@{\extracolsep{\fill}}lrrrrrrrl@{}}
\hline
$$A$ (\text{dimensionless})$& \multicolumn{1}{c}{$\rho(0)$$\times 10^{15}$\ [${g/cm^{3}}$]} & \multicolumn{1}{c}{$\rho(R)$$\times 10^{14}$\ [${g/cm^{3}}$]} & \multicolumn{1}{c}{$p_{r}(0)$$\times 10^{35}$\ [${dyne/cm^{2}}$]} & \multicolumn{1}{c}{$\Gamma$} & \multicolumn{1}{c}{$\Gamma_{\text{crit}}$}  \\
\hline
$0.25$& $1.432513188$&$8.270618749$ & $2.148769784$ & $2.333333334$ &$1.592456934$  \\
$0.20$& $1.196390717$&$7.460806943$ & $1.316029791$ & $2.550505048$ & $1.560467811$   \\
$0.15$& $0.891099759$&$0.617637224$ & $0.623076983$ & $3.079365076$ & $1.513926893$\\
\hline
\end{tabular*}
\end{table*}

\begin{table*}
\caption{The mass and compactness factor for a fixed radius $R=9.5\ [{km}]$ and for different values of the parameter $A$. }
\label{table3}
\begin{tabular*}{\textwidth}{@{\extracolsep{\fill}}lrrrrrrrl@{}}
\hline
$$A$ (\text{dimensionless})$& \multicolumn{1}{c}{$M$\ [${km}$]} & \multicolumn{1}{c}{$M_{\odot}$} & \multicolumn{1}{c}{$u\equiv\frac{M}{R}$}  \\
\hline
$0.25$& $2.720797812$&$1.846862484$ & $0.28639976$  \\
$0.20$& $2.384912024$&$1.618865072$ & $0.25104337$    \\
$0.15$& $1.896232379$&$1.287152036$ & $0.19960341$\\
\hline
\end{tabular*}
\end{table*}

\section{Equilibrium and Stability Analysis}\label{sec6}
\begin{figure*}[ht]
\centering
\includegraphics[width=0.32\textwidth]{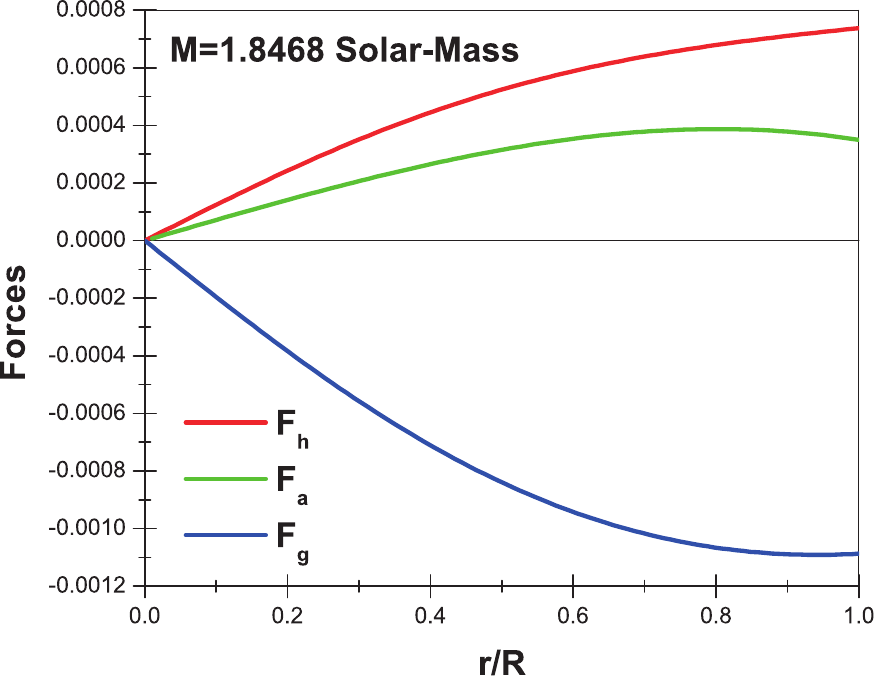}     \
\includegraphics[width=0.32\textwidth]{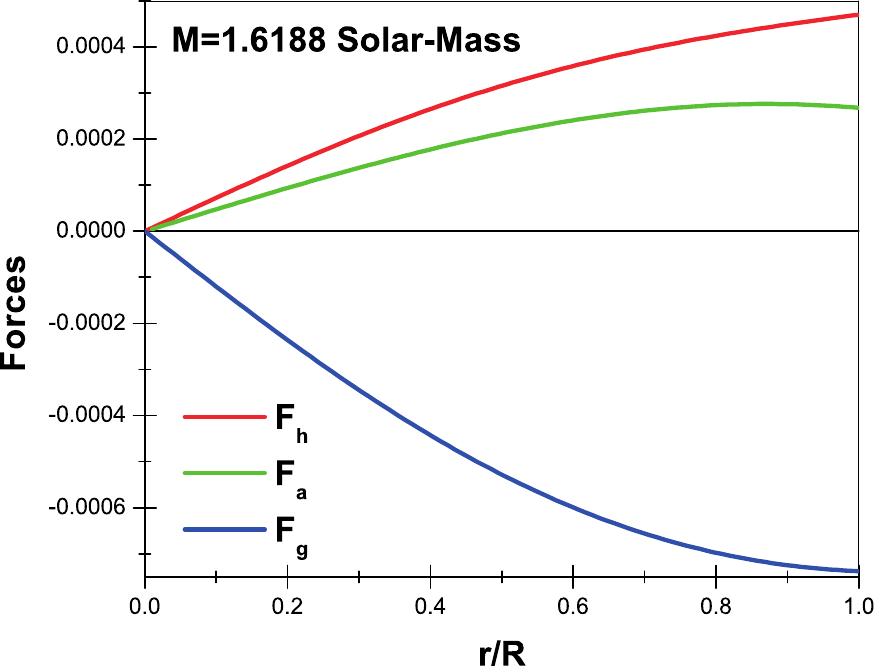}      \
\includegraphics[width=0.32\textwidth]{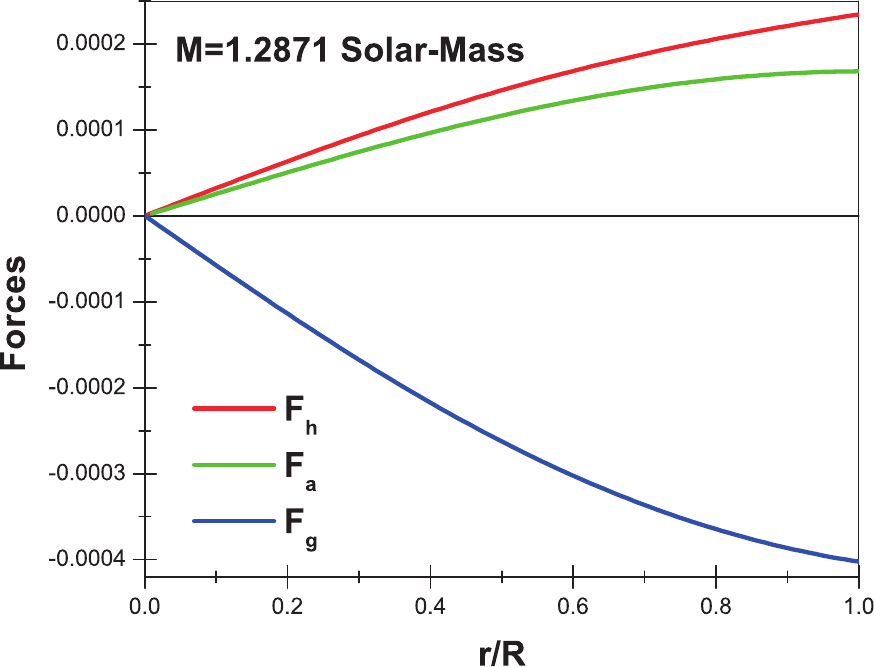}     \
\caption{
Figures show forces (hydrostatic, $F_h$, anisotropic, $F_a$, and gravitational, $F_g$) for three different values of the mass $M$ versus the normalized radial coordinate $r/R$. 
{\bf Left Panel}: Forces for $M=1.8468$ $M_{\odot}$ against $r/R$.  
{\bf Middle Panel}: Forces for $M=1.6188$ $M_{\odot}$ against $r/R$.
{\bf Right Panel}: Forces for $M=1.2871$ $M_{\odot}$ against $r/R$.
All these plots have been built by using $R=9.5\ [km]$ and different values of the parameters $A$, $B$, $C$ and $F$ mentioned in table \ref{table1}. It should be noted that, as we are working in relativistic geometrized units, the vertical axis has units of $[km^{-3}]$ for all panels.
}
\label{fig4}
\end{figure*}

\begin{figure*}[ht]
\centering
\includegraphics[width=0.32\textwidth]{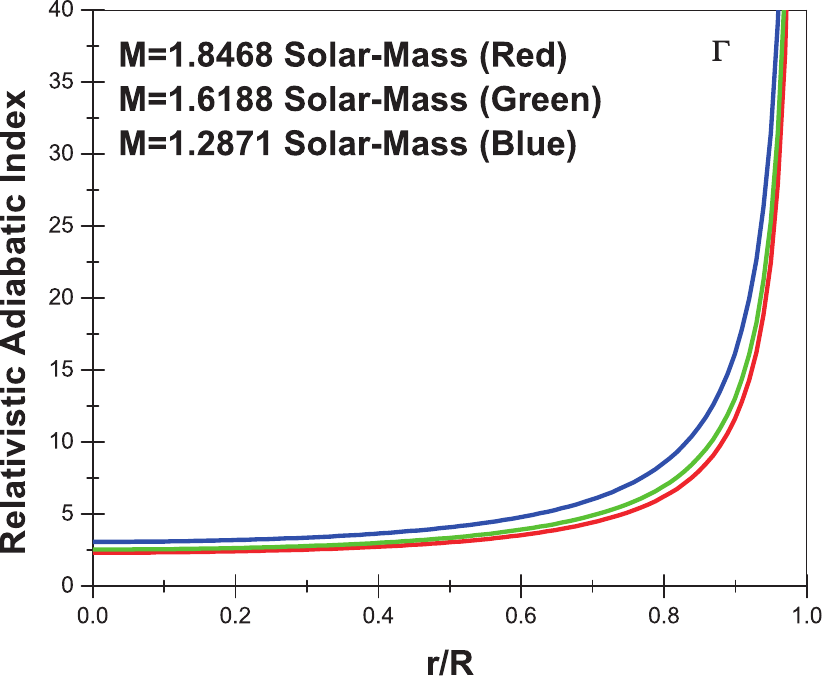}     \
\includegraphics[width=0.32\textwidth]{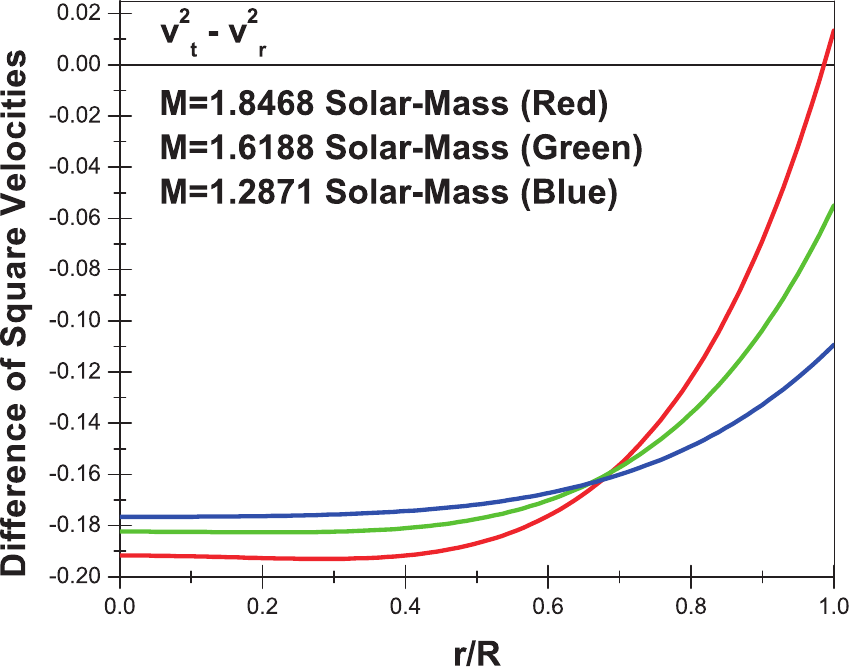}      \
\includegraphics[width=0.32\textwidth]{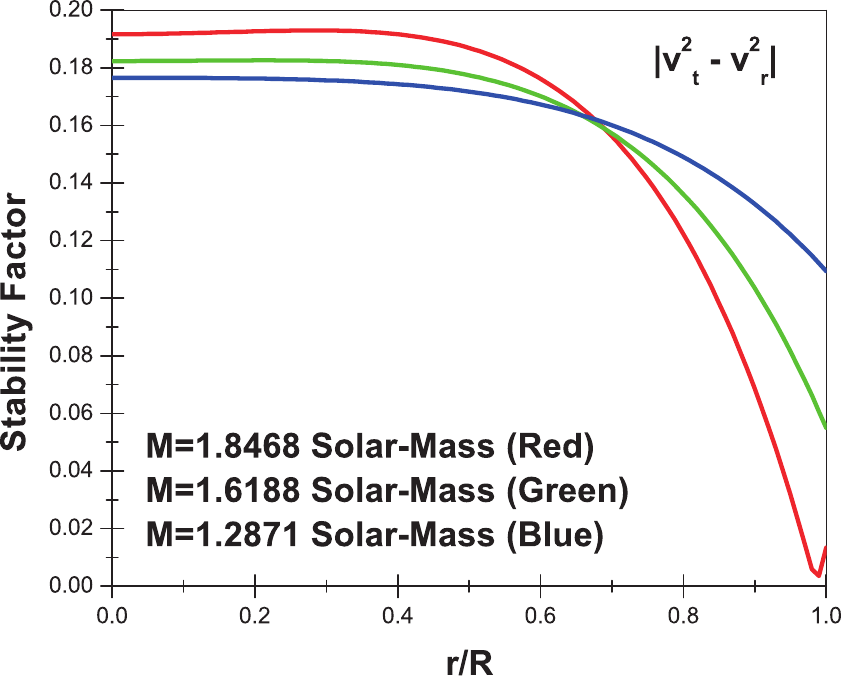}     \
\caption{
Figures show the relativistic adiabatic index, $\Gamma$, the difference of square velocities, $v_t^2 - v_r^2$, and the stability factor,  $|v_t^2 - v_r^2|$, for three different values of the mass $M$ versus the normalized radial coordinate $r/R$. 
{\bf Left Panel}: Relativistic adiabatic index against $r/R$.  
{\bf Middle Panel}: Difference of square velocities against $r/R$.
{\bf Right Panel}: Stability factor against $r/R$.
All these plots have been built by using $R=9.5\ [km]$ and different values of the parameters $A$, $B$, $C$ and $F$ mentioned in table \ref{table1}. It should be noted that, as we are working in relativistic geometrized units, the vertical axis is dimensionless for all panels.
}
\label{fig5}
\end{figure*}

Given that the system is under the effects of the gravitational, hydrostatic, and anisotropic forces, it is necessary to check if the structure is in hydrostatic balance. To verify this, all forces present in the configuration must meet 
\begin{equation}\label{eq43}
F_{h}+F_{a}+F_{g}=0,    
\end{equation}
where $F_{h}$, $F_{a}$ and $F_{g}$ stand for the hydrostatic, anisotropic and gravitational forces, respectively. The explicit form of these forces are
\begin{align}\label{eq44}
\begin{split}
F_{g} & =-\frac{d\nu}{dr}\left(\rho+p_{r}\right), 
\\
F_{a} & = \frac{2}{r}\left(p_{t}-p_{r}\right)  
\\ 
F_{h} & =-\frac{dp_{r}}{dr}.
\end{split}
\end{align}
The Eq. (\ref{eq43}) can be seen as a generalized Tolman-Oppenheimer-Volkoff equation driven the hydrostatic balance of compact anisotropic fluid spheres. As was pointed out earlier, a positive anisotropy factor $\Delta$ introduces a force repulsive in nature. This repulsive force helps to counteract the gravitational gradient produced by the gravitational force $F_{g}$. The presence of this anisotropic force repulsive in nature avoids the gravitational collapse of the structure onto a point singularity. In Fig. \ref{fig4} are depicted the balance of the system for all cases. As it is appreciated in all cases, the system is in equilibrium under the mentioned forces. However, it remains to be determined whether the equilibrium to which the system is subjected is stable or not. In this concern, talk about stability is rather heuristic; one has an idea of how the system behaves under the presence of local anisotropies, but in the realm, one can not measure the stability of the system in a closed-form. At the theoretical level, there are many ways to check the stability of the configuration under the presence of anisotropies into the matter distribution. The most common are via the relativistic adiabatic index $\Gamma$ and by means of the subliminal speed of the pressure waves \cite{r52}. In comparison with the Newtonian scenario for an isotropic fluid distribution the stability condition is $\Gamma>4/3$ \cite{r33,r41}. Nevertheless, in the relativistic case with the inclusion of local anisotropies, the requirements are completely different. So, in the anisotropic relativistic case, the adiabatic index $\Gamma$ should satisfy \cite{r42,r43}
\begin{equation}\label{eq45}
\Gamma>\frac{4}{3}+\left[\frac{1}{3}\kappa\frac{\rho_{0}p_{r0}}{|p^{\prime}_{r0}|}r+\frac{4}{3}\frac{\left(p_{t0}-p_{r0}\right)}{|p^{\prime}_{r0}|r}\right]_{max}    
\end{equation}
where $\rho_{0}$, $p_{r0}$ and $p_{t0}$ are the initial density, radial and tangential pressure when the fluid is in static equilibrium. The terms inside the brackets in (\ref{eq45}) represent the relativistic corrections and the contributions coming from the local anisotropies. However, as was pointed out by Chandrasekhar \cite{r75,r76} relativistic correction to the adiabatic index $\Gamma$ could, in principle, introduce some instabilities within the stellar interior. To overcome this issue, recently, Moustakidis \cite{r77} proposed a more strict condition on the adiabatic index $\Gamma$. In \cite{r77} was found a critical value for the adiabatic
index $\Gamma_{\text{crit}}$.  To have a stable structure, this critical value depends on the amplitude of the Lagrangian
displacement from equilibrium and the compactness factor $u\equiv M/R$. The amplitude of the Lagrangian displacement is characterized by the parameter $\xi$, so taking particular a form of this parameter the critical relativistic adiabatic index is given by
\begin{equation}\label{eq46}
\Gamma_{\text{crit}}=\frac{4}{3}+\frac{19}{21}u,    
\end{equation}
where the stability condition becomes  $\Gamma\geq \Gamma_{\text{crit}}$,
where $\Gamma$ is computed from \cite{r39}
\begin{equation}\label{eq47}
\Gamma=\frac{\rho+p_{r}}{p_{r}}\frac{dp_{r}}{d\rho}.    
\end{equation}            
From the left panel of Fig. \ref{fig5} it is observed that the system is stable from the relativistic adiabatic index point of view. What is more in table \ref{table2} are shown the numerical values for the relativistic adiabatic index $\Gamma$ and its critical values $\Gamma_{\text{crit}}$ at $r=0$. As can be seen $\Gamma >\Gamma_{\text{crit}}$ for all cases. Furthermore, the system becomes more stable for increasing $A$, because of its adiabatic index $\Gamma$ increases at $r=0$. We have also checked the stability of the compact object by using the speed criteria. Based on Herrera's cracking concept \cite{r44} Abreu and his collaborators \cite{r52} determined a form to analyze the presence of stable/unstable regions within the stellar interior when local anisotropies are there.
In short Abreu and coworkers  mechanism is built on the  subliminal speed of pressure waves as follows
\begin{equation}\label{eq48}
\frac{\delta{\Delta}}{\delta{\rho}}\sim \frac{\delta\left({p}_{t}-{p}_{r}\right)}{\delta{\rho}} \sim \frac{\delta{p}_{t}}{\delta{\rho}}-\frac{\delta{p}_{r}}{\delta{\rho}}\sim v^{2}_{t}-v^{2}_r.    
\end{equation}
Next, from causality condition one has $0\leq v^{2}_{r}\leq 1$ and $0\leq v^{2}_{t}\leq 1$ which implies $0\leq |v^{2}_{t}-v^{2}_{r}|\leq 1$. 
Explicitly it reads
\begin{equation}
\begin{split}
     \label{eq49}
   -1\leq v^{2}_{t}-v^{2}_{r}\leq 1  &\\ = \left\{
	       \begin{array}{ll}
		   -1\leq v^{2}_{t}-v^{2}_{r}\leq 0~~ & \mathrm{Potentially\ stable\ }  \\
		 0< v^{2}_{t}-v^{2}_{r}\leq 1 ~~ & \mathrm{Potentially\ unstable}
	       \end{array}
	        \right\}.
	        \end{split}
	    \end{equation}
Thus, the main idea behind Abreu's criterion is that if the subliminal tangential speed $v^{2}_{t}$ dominates the subliminal radial speed $v^{2}_{r}$; then this could potentially result in cracking instabilities. This can be easily verified by performing a graphical analysis. From the middle and right panels of Fig. \ref{fig5} it is clear that the compact object is stable for the cases corresponding to $A=0.15$ and $A=0.2$ (blue and green curves), but contains unstable regions in the case $A=0.25$ which corresponds to the highest mass value. These unstable regions are shown in the middle panel, where the quantity $v^{2}_{t}-v^{2}_{r}$ changes in sign (red curve). The right panel also confirms this in Fig. \ref{fig3} where the tangential subliminal speed $v^{2}_{t}$ overcomes the radial one $v^{2}_{r}$ approximately at $r/R=0.98$. As with the adiabatic index, the object exhibits greater stability as the mass decreases. Regarding the stability factor $|v^{2}_{t}-v^{2}_{r}|$ at $r/R=0.98$ the red curve change its direction. Again this verifies the present of cracking within the stellar interior. Despite all, one can conclude that the system is stable under local anisotropies disturbs.           

\section{The Generating Function}\label{sec7}
It has been demonstrated that all the spherically symmetric and
static isotropic solutions of the Einstein’s field equations
can be generated by choosing a single monotone function subject to the boundary conditions \cite{lake}. In more widely context Herrera et. al \cite{herrera} extended the previous work to include a more realistic and complete component \i.e, local anisotropies into the matter distribution. They concluded that all the spherically symmetric static anisotropic solutions of the Einstein’s field equations can be generated
from two generating functions $\zeta(r)$ and
$\Pi(r)$. The generator $\zeta(r)$ linked with the metric potential
$e^{\nu}$ and other with the negative of pressure anisotropy $\Delta(r)$. These generators are defined via
\begin{eqnarray} \label{eq50}
e^{2\nu(r)}&=&\text{Exp}\left[\int\left(2\zeta(r)-\frac{2}{r}\right)dr\right] \\ \label{eq51}
\Pi(r)&=&p_{r}(r)-p_{t}(r)=-\Delta(r).
\end{eqnarray}
So, from (\ref{eq50}) solving for the generator $\zeta(r)$ one gets
\begin{equation}\label{eq52}
\zeta(r)=\nu^{\prime}(r)+\frac{1}{r},
\end{equation}
where by using (\ref{eq30}) the generator $\zeta(r)$ becomes
\begin{equation}\label{eq53}
\begin{split}
\zeta(r)=\frac{1}{r}+\frac{r}{4C\left(Cr^{2}+1\right)\left(Cr^{2}+3\right)}\bigg[BC^{3}r^{6}+5BC^{2}r^{4}&\\
+4AC^{3}r^{2}+9BCr^{2}+12AC^{2}+5B\bigg].    
\end{split}
\end{equation}
Now, from expressions (\ref{eq23}) and (\ref{eq24}) the anisotropy factor $\Delta(r)$ associated with the model is given by 
\begin{equation}\label{eq55}
\begin{split}
\Delta(r)=\frac{-1}{16C^{2}\left(Cr^{2}+1\right)^{3}\left(Cr^{2}+3\right)^{2}}\bigg[B^{2}C^{7}r^{16}&\\+9B^{2}C^{6}r^{14}+8ABC^{7}r^{12}+33B^{2}C^{5}r^{12}&\\
+56ABC^{6}r^{10}+16A^{2}C^{7}r^{8}+64BC^{6}r^{10}&\\
+32AC^{7}r^{8}+57B^{2}C^{4}r^{10}+128ABC^{5}r^{8}&\\
+80A^{2}C^{6}r^{6}+144BC^{5}r^{8}+384AC^{6}r^{6}&\\
31B^{2}C^{3}r^{8}+64ABC^{4}r^{6}+32C^{6}r^{6}&\\
+48A^{2}C^{5}r^{4}+64BC^{4}r^{6}+1504AC^{5}r^{4}&\\ -41B^{2}C^{2}r^{6}-136ABC^{3}r^{4}+224C^{5}r^{4}&\\
-144A^{2}C^{2}r^{2}-200BC^{3}r^{4}+2112AC^{4}r^{2}&\\
-65B^{2}Cr^{4}-120ABC^{2}r^{2}+480C^{4}r^{2}&\\
-320BC^{2}r^{2}+576AC^{3}-25B^{2}r^{2}&\\+288C^{3}-144BC
\bigg].
\end{split}
\end{equation}
So, the second generator $\Pi(r)$ is 
\begin{equation}\label{eq54}
\begin{split}
\Pi(r)=-\Delta(r)=\frac{1}{16C^{2}\left(Cr^{2}+1\right)^{3}\left(Cr^{2}+3\right)^{2}}\bigg[B^{2}C^{7}r^{16}&\\+9B^{2}C^{6}r^{14}+8ABC^{7}r^{12}+33B^{2}C^{5}r^{12}&\\
+56ABC^{6}r^{10}+16A^{2}C^{7}r^{8}+64BC^{6}r^{10}&\\
+32AC^{7}r^{8}+57B^{2}C^{4}r^{10}+128ABC^{5}r^{8}&\\
+80A^{2}C^{6}r^{6}+144BC^{5}r^{8}+384AC^{6}r^{6}&\\
31B^{2}C^{3}r^{8}+64ABC^{4}r^{6}+32C^{6}r^{6}&\\
+48A^{2}C^{5}r^{4}+64BC^{4}r^{6}+1504AC^{5}r^{4}&\\ -41B^{2}C^{2}r^{6}-136ABC^{3}r^{4}+224C^{5}r^{4}&\\
-144A^{2}C^{2}r^{2}-200BC^{3}r^{4}+2112AC^{4}r^{2}&\\
-65B^{2}Cr^{4}-120ABC^{2}r^{2}+480C^{4}r^{2}&\\
-320BC^{2}r^{2}+576AC^{3}-25B^{2}r^{2}&\\+288C^{3}-144BC
\bigg].
\end{split}
\end{equation}
\section{Concluding Remarks}\label{sec8}

In this work, it has been reported a relativistic anisotropic fluid sphere in the arena of general relativity. This solution of the Einstein field equations is representing compact objects such as quark stars admixed with dark energy. The inclusion of dark energy was introduced via the so-called modified generalized Chaplygin equation of state. To close the problem from the mathematical point of view, we have imposed a suitable form of the metric potential $e^{2\lambda(r)}$ and the mentioned equation of state. Regarding the former, it has been previously employed in \cite{r30}. The particularity of this choice is that it is free from mathematical singularities and leads to a well-defined density $\rho(r)$ and mass function $m(r)$. Regarding the thermodynamic relation given by Eq. (\ref{chaplygin}), this corresponds to a non-linear link between the radial pressure $p_{r}(r)$ and density $\rho(r)$ by governing the micro-physics process of the imperfect matter distribution in the stellar interior. This constrains between the main physical variables of the system depends on to parameters, namely, $A$ and $B$, both are strictly positive numbers, and $A$ is restricted to belongs to $(0,\frac{1}{3})$. Moreover, to simplify the mathematical inconvenient, we have employed as usual the Durgapal-Barneerji transformation \cite{r71}. The same procedure has been used earlier by other authors \cite{r5,r6}.

Once the complete geometry (\ref{eq26}) and (\ref{eq30})  describing the inner space-time is obtained we have computed the full set of thermodynamic observables $\{\rho,p_{r},p_{t}\}$ given by Eqs. (\ref{eq20}), (\ref{eq23}) and (\ref{eq24}) respectively. After that, to close the problem, the matching condition mechanism via Israel-Darmois junction conditions was employed to determine all the constant parameters $\{B,C,F\}$, and the total mass $M$ that characterize the model. Due to we dealt with an uncharged anisotropic matter distribution, the compact object is surrounded by a vacuum space-time described by the well known Schwarzschild solution. Besides, to mimic a realistic collapsed structure \cite{r31}, we have taken the radius of the configuration to be $9.5\ [km]$. It is found that moving the constant parameter $A$ from $0.15$ to $0.25$ the total mass $M$ increases progressively from $1.2871 M_{\odot}$ to $1.8468 M_{\odot}$ (for more details see table \ref{table3}). These values are in complete agreement with those reported in \cite{r68}.
Furthermore, as can be seen from Table \ref{table2}, in all cases the central density $\rho(0)$ overcome the nuclear density saturation ($\rho\sim 2.4\times 10^{14}\ [g/cm^{3}]$). This in conjunction with the high central pressure suggest that the core of the compact star is formed by the so--called strange matter \i.e, matter conformed by quarks up, down and strange. Then this toy model could represents quark (strange) stars. What is more the assumed radius and the obtained masses are to close to the range hypothesized in previous works for strange stars \cite{r78}. Commonly, it is assumed that quarks are made of free fermions constrained within a bag with a vacuum
pressure that keeps the particles within this bag. This universal pressure is denoted by $B_{g}$ and it is known as the bag constant. The EoS driven this kind of matter is the well--known MIT EoS expressed by \cite{r12}
\begin{equation}
p=\frac{1}{3}\left(\rho-4B_{g}\right).   
\end{equation}
The studies by Farhi and Jafee \cite{new1} and Alcock et al \cite{r79} had shown that for a stable strange quark matter the value of the bag constant should be $B_{g}\sim 55-75$ $MeV/fm^{3}$. However, the datasets of CERN-SPS and RHIC22 \cite{new2} show that a wide range of bag constant are permissible. In this regard in \cite{tello} (and references contained therein) was shown that in the framework of Einstein theory the usage of the MIT EoS admits a wide range for the bag constant $B_{g}$ obtaining well behaved compact structures. From the phenomenological point of view several authors have proposed that a real EoS $p=p(\rho)$ relating the main thermodynamic variables of the fluid distribution inside the stellar interior should be well approximated by a linear function of the energy density $\rho$ \cite{dey,harko,gondek}. In this sense the MIT EoS presents such morphology but it is restricted to describe cold and massless quark matter only. So, a question arise: It is possible to obtain a quark star model, starting from a completely general linear EoS? in the affirmative case, are these objects stable? \cite{new3}. Despite we have used a highly non--linear EoS to link the the radial pressure and density, the resulting numerical data matches previous results in the same field. However, it must be taken into account that the imposition of an EoS (regardless of the form it has) is limited by three restrictions \cite{new4}: i) laboratory measurements of nuclear properties and reactions, ii) theoretical ab-initio calculations,
and iii) observations in astronomy. As said before, we have imposed the modified generalized Chaplygin equation, which is an alternative to explain the presence of the dark energy filling the whole Universe \cite{r17,r18}. So, regarding the previous comments about the numerical data and this latest ingredient, we can conclude that the solution reported in this work could represent quark stars admixed with dark energy. What is more, massive quarks (strange quarks) inspires the dark matter production from the Big Bang, representing the ground state of baryonic matter \cite{new5,new6,new7}. So, the compact object constitutes a completely dark structure composed by dark matter and dark energy, where the interaction between dark energy and dark matter problem related with the coincidence one, which is based on the fact that the ratio between dark energy and
dark matter energy densities is nowadays of the order $73/23$, while at the Planck time this ratio was of the order $10^{-95}$, remains open. Nevertheless, this problem can be faced on the background of isotropic homogeneous cosmological model by assuming a flat space--time \cite{new8}. Then, if a compact configuration driven by dark components exits it could be possible to tackle the mentioned problem from a different perspective to the cosmological one. In this regard, the existence of dark stars has been supported by the study about accretion of dark matter particles \cite{new9,new10}. To back up the previous discussion, we have checked the feasibility of our model by performing an exhaustive physical and mathematical analysis. This analysis concerns the basic and general requirements that all solutions of the Einstein fields equations must meet in order to be an admissible model. Those requirements are \cite{r34,r51}
\begin{enumerate}
    \item The geometry describing the inner manifold should be free from mathematical and physical singularities.
    \item The main thermodynamic variables should be monotone decreasing functions with increasing radius and strictly positive functions everywhere inside the object.
    \item The subliminal sound speed in the principal directions \i.e, $v^{2}_{r}$ and $v^{2}_{t}$ of the pressure waves must the less or equal than the speed of light to preserve causality condition.
    \item The energy-momentum tensor representing the mater distribution should satisfy some constraints to have a well-defined material content driven the stellar interior.
    \item The compact structure must remain in equilibrium under the presence of the hydrostatic $F_{h}$, gravitational $F_{g}$, and anisotropic $F_{a}$ forces.
    \item The hydrostatic balance should be stable. To check the stability of the configuration under radial perturbations introduced by the anisotropies, one can use, for instance, the relativistic adiabatic index or Abreu's criterion.
\end{enumerate}
The first two points have been extensively discussed before. Moreover, they are supported by Figs. \ref{fig1} and \ref{fig2} where it is clear that the metric potentials $e^{2\nu}$ and $e^{2\lambda}$ are regular everywhere and $\rho$, $p_{r}$ and $p_{t}$ are positive defined at all point within the star and monotone decreasing functions from the center $r=0$ towards the surface of the structure. Respect to the remaining points Figs. \ref{fig3}, \ref{fig4} and \ref{fig5} corroborate that the energy-momentum tensor representing the imperfect matter distribution is strictly positive, the subliminal sound velocities of the pressure waves in the along the main direction of the fluid sphere respect causality condition and that the system is under hydrostatic balance being it stable from the relativistic adiabatic index and Abreu's criterion point of view.

Nevertheless, in considering the latest analysis, as the constant parameter $A$ increases, the system presents unstable regions. This is because, at some point inside the star, the transverse sound speed overcomes the radial one, which means that the system is suffering cracking \cite{r44}. That is the main reason why the maximum value considered for $A$ is $0.25$. Notwithstanding, the role played by the anisotropy is fundamental in the balance and stability of the system. In the present study, the system experiences a positive anisotropy factor $\Delta$ introducing into the configuration a force repulsive in nature that helps to counteract the gravitational gradient, what is more with a positive anisotropy factor it is possible to build more compact objects \cite{r45} than one is dealing with isotropic fluids.  
In general terms, we can conclude that the reported model could serve to represent realistic compact objects in the arena of general relativity, which matter distribution is driven by an imperfect fluid. However, it should be considered that the arbitrary parameters that characterize the model, namely $\{A, B, C, F\}$ must be delimited, in order to reproduce a numerical data that conforms to the analyzes reported by numerical simulations or phenomenological studies. In this case, the analytical toy model represents a very good approximation to describe structures as complex as those mentioned above.

\section*{Acknowledgements}
F. Tello-Ortiz thanks the financial support by the CONICYT PFCHA/DOCTORADO-NACIONAL/2019-21190856 and projects ANT-1856 and SEM 18-02 at the Universidad de Antofagasta, Chile, and grant Fondecyt No. $1161192$, Chile and TRC project-BFP/RGP/CBS/19/099 of the Sultanate of Oman. The author \'A. R. acknowledges DI-VRIEA for financial support through Proyecto Postdoctorado 2019 VRIEA-PUCV.

\end{document}